\documentclass[preprint,showpacs,preprintnumbers,amsmath,amssymb]{revtex4}

\usepackage{graphicx}
\usepackage{dcolumn}
\usepackage{bm}

\begin{document}

\title {Magnetization Dynamics, Gyromagnetic Relation, and Inertial Effects}
\author{J.-E.
Wegrowe; M.-C. Ciornei} \affiliation{Ecole Polytechnique, LSI, CNRS and
CEA/DSM/IRAMIS, Palaiseau F-91128, France.}

\date{\today}

\begin{abstract}
The gyromagnetic relation - i.e. the proportionality between the angular momentum $\vec L$ (defined by an inertial tensor) and the magnetization $\vec M$ - is evidence of the intimate connections between the magnetic properties and the inertial properties of ferromagnetic bodies. However, inertia is absent from the dynamics of a magnetic dipole (the Landau-Lifshitz equation, the Gilbert equation and the Bloch equation contain only the first derivative of the magnetization with respect to time).  In order to investigate this paradoxical situation, the lagrangian approach (proposed originally  by T. H. Gilbert) is revisited keeping an arbitrary nonzero inertial tensor. A dynamic equation generalized to the inertial regime is obtained. It is shown how both the usual gyromagnetic relation and the well-known Landau-Lifshitz-Gilbert equation are recovered at the kinetic limit, i.e. for time scales above the relaxation time $\tau$ of the angular momentum. 
\end{abstract}


\maketitle


The analogy between the dynamics of the magnetization in a magnetic filed on one hand, and 
the dynamics of a symmetrical spinning top in a gravitational field on the other hand, is often exploited in 
introductory courses on magnetism. The precession effect (i.e. the rotation of the extremity of a vector of constant modulus) is indeed 
easy to observe on a spinning top, while it is difficult to see with 
a 
ferromagnet because it would require observations at sub-nanosecond 
time scales. \cite{Stoehr} However, the analogy seems to be incomplete because the dynamics of the symmetric spinning top 
implies inertial effects 
(e.g. nutation) while for a uniformly 
magnetized body, the 
dynamics of the magnetization is described by the time variation of 
the magnetization $d\vec M/dt$ (i.e. the velocity) and does not include the second derivative $d^2\vec 
M/dt^2$ (i.e. the acceleration). \cite{Doering} In other terms, there is no inertia in the 
dynamic equation. The aim of this paper is to push the analogy to 
its logical end with the introduction of inertia  \cite{inertial} in 
the dynamics of uniform magnetization within the Lagrangian formalism.

The precession of a uniform magnetic moment $ \vec M = M_s \vec e_3$ 
($M_s$ is the magnetization at saturation and $\vec e_3$ the radial 
unit vector) under an effective magnetic field 
$\vec H$ is often presented as a consequence of the gyromagnetic 
relation $\vec M = \gamma \vec L$ that links the 
magnetization to the angular 
momentum $\vec L$.  The constant $\gamma$ is the 
gyromagnetic ratio. The gyromagnetic relation and 
the value of the constant $\gamma = q/(2m)$ can be justified in a basic {\it  atomic} model of an electron of charge 
$q$ and mass $m$ orbiting around a nucleus. This well-known model (see section IV below) constitutes the hypothesis of the Amp\`ere molecular currents, 
validated by Einstein and de Haas in their famous experiments of 1915 
- 1916. \cite{Einstein,Frenkel} In the general case, with both spin and orbital contributions in condensed material, the gyromagnetic ratio writes  $\gamma = g \, q/(2m)$ where the $g$ factor accounts for  the fact that the electron in a ferromagnet is a complex quasi particle. \cite{Ohanian,Stoehr}

Using the gyromagnetic relation, 
the application of Newton's second law $d \vec L/dt = \vec M 
\times \vec H$ leads directly to the precession equation $d\vec M/dt = 
\gamma 
\vec M \times \vec H$. However the application of the Newton's law to a rigid rotating body (typically the spinning top in a gravitational field), leads to a more complex gyroscopic equation that contains inertial terms. As will be shown below, the gyromagnetic relation also imposes inertia for the  dynamics of the magnetization. This paradoxical situation can be clarified by re-introducing the inertia in the equation of the magnetization and explicitly going to the kinetic limit. 
 
It is first useful to come back to the short history of the dynamic equation of the magnetization, especially with the introduction of  the dissipation, since the precession equation  $d\vec M/dt = 
\gamma 
\vec M \times \vec H$ cannot 
account for the rapid relaxation toward the 
equilibrium state of the magnetization (typically after a couple of precession cycles, i.e. after some nanoseconds, in usual ferromagnets).  In 1935 Landau and Lifshitz proposed an equation for the 
dynamics of
the magnetization that takes into account both the 
precession and the relaxation along the magnetic field: 
$d\vec{M} /dt = \tilde \gamma \vec M \times 
\vec{H} \, + \,  h' \vec{M} \times (\vec{M}
\times \vec{H})$, where $h'$ is a damping term 
(defined 
below) and $\tilde \gamma = \gamma$.  \cite{Landau} The 
basic argument used to derive the equation was to keep the modulus of the 
magnetization constant. The derivative $d\vec 
M/dt$ is hence perpendicular to the vector $\vec M$.

Two decades later, after the development of ferromagnetic resonance (FMR)
experiments \cite{Bloembergen} and motivated by the observation of 
systematic deviations from the 
above equation for high damping,  T. 
L. Gilbert derived the equation that bears
his name using a Lagrangian formalism.  
\cite{GilbertPhD,Gilbert}  The dynamics of the magnetization is then 
described 
by the equation
\begin{equation}
\frac{d\vec{M}}{dt} = \gamma \vec{M} \times \left ( \vec{H} - 
\eta
\frac{d\vec{M}}{ dt} \right )
\label{Gilbert}
\end{equation}
with the introduction of the damping coefficient $\eta$.

The Landau - Lifshitz equation and the Gilbert equation
are equivalent \cite{Miltat} provided that $h' = \frac{\alpha
\gamma}{1 + \alpha^{2}}$ and $\tilde \gamma = \frac{\gamma}{1 +
\alpha^{2}}$ where $\alpha = \gamma \eta M_{s}$ is the dimensionless
Gilbert damping (note that  $\tilde \gamma \ne \gamma$: this was the decisive improvement brought by Gilbert to the Landau-Lifshitz proposition). 

  In line with previous works performed by W. D\"oring, \cite{Doering} Gilbert introduced the Lagrangian of a uniform ferromagnet with a
kinetic energy $\mathcal{T}= \vec{L} \vec{L}: \bar{\bar{I}}^{-1}/2$, where $\bar{\bar{I}}$ is the inertial tensor. 
He then chose an ad-hoc tensor of inertia in such a way that the
inertial terms disappear from the dynamic equation (i.e. such that the Landau-Lifshitz equation is recovered at the low damping limit).  To do that, a
sufficient condition is to set to zero the two first principal moments
of inertia $I_{1}=I_{2}=0$ (but keeping a non-zero kinetic energy: $I_3 \ne 0$).  As pointed out by Gilbert himself
\cite{Gilbert} this puzzling condition does not seem to correspond to any
realistic mechanical system (see footnote 7 : "{\it I was unable to conceive of a physical object with an inertial tensor of this kind}"). 
 In the subsequent report about Gilbert's derivation, the ad-hoc and puzzling condition
$I_{1}=I_{2}=0$ is explicitly stated despite its problematic character. In his presentation of the Gilbert equation published in 1960 in the {\it American Journal of Physics}, \cite{Brown} Brown wrote "{\it We treat the rotating moment system as a symmetric top, with 
principal moments of inertia $A= B = 0$, $C > 0$. For a top made of 
classical mass particles, $ A = B = 0$ implies $C= 0$; but this top is 
not made of classical mass particles.}" In our notation $A  \equiv 
I_1$, $  B  \equiv I_2$, $C \equiv I_3$. 
In the reference textbook of Morrish \cite{Morrish} (edited from 1965 to 2002), we can read:
 "{\it A Lagrangian function, $\mathcal{L}$, consistent with the 
accepted equation of motion (equation (10-3.2)) can be obtained by 
considering the magnetic system as a classical top with principal 
moments of inertia $(0,0,C)$...}". In our notation $C \equiv I_3$, and the equation (10-3.2) is 
$d\vec M/dt = \gamma \vec M \times \vec H + damping$. Accordingly, the mechanical approach is not presented as a realistic physical model (as it should, according to the gyromagnetic relation), but seems to be  introduced as a pedagogical  analogy of an unspecified non-classical theory, that would give a physical interpretation to the puzzling Gilbert's condition. Indeed, this strange condition is presented as a specific property of the magnetic moments that would be due to the fact that "{ \it this top is not made of classical mass particles" }. 

This is probably the reason why, after more than half a century of intensive use of the Gilbert's
equation, the full derivation following Gilbert's approach - with the
complete set of principal moments of inertia (i.e. without ad-hoc
assumption) - has not been proposed (see e.g. \cite{Ricci,Miltat} for
recent presentations of the Gilbert's derivation).  However, as
will be shown below, this derivation can be performed at an elementary
level, as a direct application of the Lagrangian formalism.  Although
straightforward, this derivation is very instructive because it shows
that the puzzling condition $I_1 = I_2 = 0$ is not necessary to obtain
the Gilbert equation. Instead, the Gilbert's condition is replaced by the necessary physical condition under which a diffusion process can be described by a non-inertial diffusion equation. This condition is the usual kinetic limit that results in the requirement that the typical measurement times should be
longer than the relaxation time $\tau$ of the momentum (here for the angular momentum $\tau = I_1 /(\eta M_s^2)$).  \cite{Rubi} In this picture, the precession with damping is simply a diffusion process in a field of force, for which the angular momentum has reached its equilibrium. This change of paradigm has two consequences. A first important consequence is that an inertial regime of uniform magnetic dipoles is expected, and should be observed at short enough time scales. Second, the classical mechanical approach is much more than a pedagogical analogy, and it could be used (beyond the gyromagnetic relation) for a deeper understanding of non-equilibrium magnetomechanics and related  processes.

\section{The mechanical analogy}
 
The mechanical model is sketched in Figure 1. A rigid
stick of length $M_s$ with one extremity fixed at the origin is described by the angles $\theta$ and $\varphi $. The stick is precessing around the vertical axis at the angular velocity $\dot \varphi$ and is spinning around its symmetry axis at the angular velocity $\dot \psi$. The phase space of this rigid rotator is
defined by the angles $\{ \theta, \varphi, \psi \} $ plus the angular momentum $\vec L$. The relation between the angular momentum 
and the angular velocity $\vec \Omega$ is $\vec L =  \bar{\bar{I}} 
\vec \Omega$ where $\bar{\bar{I}}$ is the inertial tensor. 

\subsection{The rotating frame}

 In the rotating frame, or body-fixed frame
$\{ \vec e_1, \vec e_2, \vec e_3 \}$, the inertial tensor 
is reduced to the principal moments of inertia $\{ I_1,I_2,I_3 \}$. 
The symmetry of revolution imposes furthermore that $I_1=I_2$:

\begin{equation}
\bar{I} = \left( \begin{array}{ccc}
 I_1 &  0 & 0 \\
                       0 & I_1 & 0 \\
                       			0 & 0 & I_3
\end{array} \right)
\label{MatrixL0}
\end{equation}

 \begin{figure} [h!]
   \begin{center}
   \begin{tabular}{c}
   \includegraphics[height=10cm]{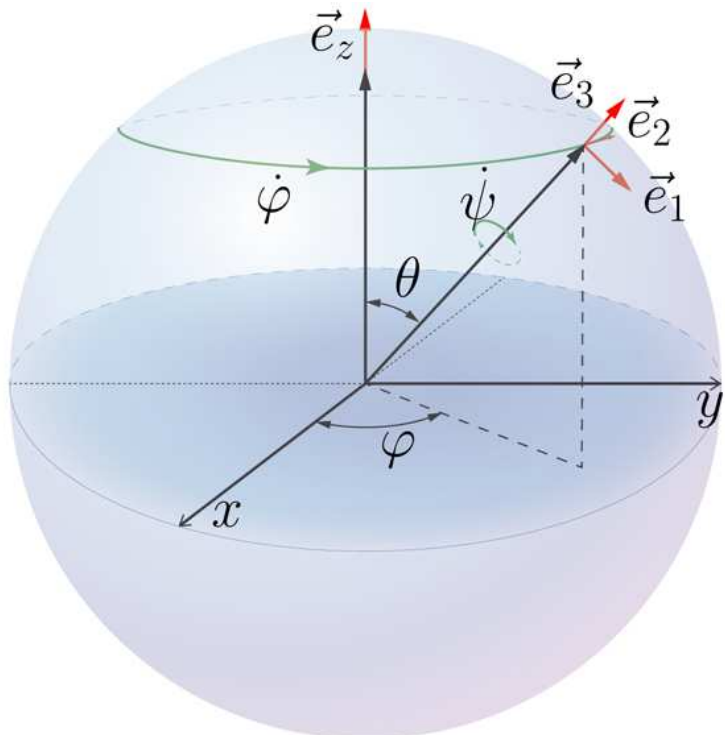}
   \end{tabular}
   \end{center}
   \caption[SpinTop]
{ \label{fig:SpinTop} Illustration of the magnetomechanical analogy of a 
spinning stick that precesses around the $z$ axis. The coordinates of the stick in the space-fixed frame are parametrized by the angles $ ( \theta,  \varphi, \psi )$ and the radius of the sphere is given by $M_s$. The body-fixed frame - denoted $\{\vec e_1,\vec e_2, \vec e_3\}$ - is spinning with the angular velocity $\dot \psi$ and is precessing around $\vec e_z$ with the angular velocity $\dot \varphi$.}
   \end{figure}

In the fixed body frame, the angular velocity reads (see Fig. 1):

\begin{equation}
\begin{array}{ccc}
 \Omega_1 & = & \dot \varphi  \, sin \theta \, sin \psi + \dot \theta 
\, cos \psi \\
 \Omega_2  &=& \dot \varphi  \, sin \theta \, cos \psi - \dot \theta 
\, sin \psi \\
  \Omega_3 & =& \dot \varphi  \, cos \theta \  + \dot  \psi
\end{array}
\label{Omega0}
\end{equation}

The kinetic equation is obtained from the 
angular velocity:  for any vector $\vec M$ of constant modulus carried with the rotating body, 
we have

\begin{equation}
\frac{d \vec M}{dt} = \vec \Omega \times \vec M 
\label{kinetic}
\end{equation}  

This equation can be inverted by cross multiplication by $\vec M$ 
and developing the double cross product. Since $\vec M = M_s \vec e_3$ we have:

\begin{equation}
\vec{\Omega} = \frac{\vec{M}}{M_{s}^{2}} \times \frac{d\vec{M}}{dt} + 
\Omega_3 \vec e_3
\label{kineticBIS}
\end{equation}

\subsection{The Lagrange equation}

Following Gilbert and D\"oring, we introduce the Lagrangian of the 
system: $$\mathcal{L} = \frac{1}{2}\left ( I_1 \left ( \Omega_1^2 + 
\Omega_2^2 \right) + I_3 \Omega_3^2 \right) - V(\theta, \varphi)$$
where $V(\theta, \varphi)$ is the ferromagnetic potential energy that defines the {\it effective magnetic field}
$\vec H = - \vec \nabla V$. The effective field $\vec H$ comprises the applied field, the anisotropy field, the dipolar field (or the demagnetizing field), the magneto-elastic contributions, etc. 

The Lagrange equations are defined by :

\begin{equation}
 \frac{d}{dt} \frac{\partial \mathcal L}{\partial \dot q_i} - 
\frac{\partial \mathcal L }{\partial q_i} + \frac{\partial \mathcal F 
}{\partial \dot q_i} = 0
\label{Lagrange}
\end{equation}

The $q_i$ refers to the three coordinates $\{\theta, \varphi, \psi 
\}$, and the components of the kinetic momentum are 
defined by the three derivatives $\frac{\partial \mathcal L}{\partial 
\dot q_i} = L_i$. The function $\mathcal F$ is the Rayleigh 
dissipative function. In a viscous environment, the Rayleigh function 
is defined by the damping coefficent $\eta$ such that $\mathcal F = 
\frac{\eta}{2} (\frac{dM}{dt})^2 = \frac{1}{2} \eta M_s^2 \left ( 
\Omega_1^2 + \Omega_2^2\right)$.

For the magnetomechanical model, the Lagrange equations read:

\begin{equation}
\begin{array}{ccc}
 \frac{d }{dt} [I_1 \dot \theta]    - I_1 \dot \varphi^2  \, sin 
\theta \, cos \theta + I_3  \dot \varphi  \, sin \theta \left (  \dot 
\varphi \, cos \theta + \dot \psi \right 
)  & =  & - \frac{\partial \mathcal F }{\partial 
\dot \theta}  - \frac{\partial V }{\partial  \theta} \\
  \frac{d }{dt} \left  [ I_1 \dot \varphi sin^2 \theta + I_3 \left( 
\dot \varphi  \, cos \theta  + \dot \psi \right) cos \theta \right 
]  \, & =  \,  & - \frac{\partial \mathcal F }{\partial \dot \varphi} - 
\frac{\partial V }{\partial  \varphi}  \\
  \frac{d }{dt}  \left  [ I_3 \left( \dot \varphi  \, cos \theta + 
\dot \psi \right) \right ] \, & = & \, - \frac{\partial \mathcal F 
}{\partial \dot \psi}  \,  = 0
  \end{array}
\label{Lagrange1}
\end{equation}
\\

The right hand side of Eqs.~(\ref{Lagrange1}) the last equation is equal to zero because there is no damping for spinning in the case of usual viscous environment. 
\cite{Condiff} The quantity $L_3 = I_3 \Omega_3$ is then a
constant of motion, and it can be written $L_3 = M_s/\gamma$ without loss of generality.

\section{Kinetic equation and Gilbert's assumptions}

It is not trivial to see how to recover the Landau-Lifshitz equation 
from Eqs.~(\ref{Lagrange1}), even at the low damping limit $\eta 
\rightarrow 0$. But it is clear that the inertial terms in the left 
hand side of Eqs.~ (\ref{Lagrange1}) are not welcome from that point of 
view and should be removed. The best way to consider the inertial 
terms is to take the kinetic Equation Eq.~ (\ref{kineticBIS}) with 
$\vec L = \bar{\bar{I}} \vec \Omega$:    

\begin{equation}
 \vec L = \frac{I_1}{M_{s}^{2}} \left  ( \vec{M} \times 
\frac{d\vec{M}}{dt} \right ) +  L_3 \vec e_3
\label{kinetic2}
\end{equation}

It is then rather immediate to see that the gyromagnetic relation 
$\vec L = \vec M / \gamma$ cannot be recovered without removing the first 
term on the right hand side. This was the great idea of Gilbert to assume that
 $I_1 = 0$.  This is indeed a sufficient condition to kill the 
inertial terms, and the gyromagnetic relation is necessarily recovered with the definition $\gamma = M_s/L_3$ of the gyromagnetic ratio.

With both assumptions, the Lagrange equations Eqs. 
(\ref{Lagrange1}) rewrite:

\begin{equation}
\begin{array}{ccc}
0 &=& - \frac{M_s}{\gamma}   \dot \varphi  \, sin \theta - 
\frac{\partial \mathcal F }{\partial \dot \theta}  - \frac{\partial V 
}{\partial  \theta} \\
  \frac{d }{dt} \left  [  \frac{M_s}{\gamma} cos \theta \right ]   & 
= & - \frac{\partial \mathcal F }{\partial \dot \varphi} - 
\frac{\partial V }{\partial  \varphi}  \\
  \frac{d }{dt}  \left  [ \frac{M_s}{\gamma} \right ]  & = & 0
\label{Lagrange2}
  \end{array}
\end{equation}



Since in the rotating frame the effective field $\vec H = - \vec \nabla V$ reads $ \{ H_1 = - 
\frac{1}{\sin(\theta)} \frac{\partial V }{\partial \varphi} ,H_2 = 
\frac{\partial V}{\partial \theta},  \} $, inserting Eq.~(\ref{kinetic}) into Eq.~(\ref{Lagrange2}) 
leads to :

\begin{equation}
 \frac{d \vec M}{dt} = \gamma \vec M \times \left (
\vec H - \eta \frac{d \vec M}{dt}  
\right)
\label{Gilbert2}
\end{equation}

This is the well-known Gilbert's equation Eq.~(\ref{Gilbert}) obtained following  the standard Lagrangian approach. \cite{GilbertPhD,Gilbert, Brown,Morrish,Ricci,Miltat}

However, the absence of inertia shows that the equation should be derived in the configuration space instead of the phase space  (this is performed e.g. in references \cite{JEW}). Indeed, the dynamics is described by the two variables $\theta$ and $\varphi$ and not in the phase space defined by the five variables $\theta$, $\varphi$ and the components of $\vec L$. Accordingly, the gyromagnetic relation is - in this approach- not necessary (nor sufficient) for the derivation of the Gilbert equation. 

\section{Beyond Gilbert's assumption}

In order to take into account the gyromagnetic relation, it is necessary to go beyond the Gilbert's ad-hoc assumption and to set $I_1 = I_2 \ne 
0$. The generalization of the Amp\`ere molecular model from a quasi-one dimensional atomic model (the electric charge $q$ of mass $m$ distributed along the circular orbit) to a more realistic three dimensional atomic model - for which the orbits form an ellipsoid of revolution (the electric charge is now distributed in three dimensions) - imposes non-vanishing inertial moments $I_1=I_2 \le I_3 $. Indeed, two parameters are necessary to take into account the amplitude of the magnetic moment ($M_s = \gamma /(\Omega_3 I_3)$) on one hand, and the anisotropy of the ferromagnetic material (with the dimensionless parameter $1-I_1/I_3$) on the other hand.

Note that for
the magnetic system, the variables $\psi$ and $\dot \psi$ are not
defined and should be removed from the model.  Let us take for the
sake of simplicity $\dot \psi = 0$. \cite{Rquedotpsi} With the relation $L_{3} =  \frac{M_s}{\gamma}$,  Eq.~(\ref{Lagrange1}) gives:

\begin{equation}
 \begin{array}{ccc}
\dot \Omega_1  &=& - \frac{\Omega_1}{\tau} + \Omega_3 \left ( 1- 
\frac{I_3}{I_1} \right )  \Omega_2 - \frac{M_s}{I_1} H_2  \\
 \dot \Omega_2 &=& - \frac{\Omega_2}{\tau}  -  \Omega_3 \left ( 1- 
\frac{I_3}{I_1} \right )  \Omega_1 + \frac{M_s}{I_1} H_1 \\
\Omega_3  & = & \frac{M_s}{\gamma I_3}
 \end{array}
\label{Lagrange4}
\end{equation}

where the typical relaxation time $\tau \equiv \frac{I_1}{\eta M_s^2}$ has been introduced. The relaxation time $ \tau$ is the typical time above which the diffusion approximation is valid, i.e. above which the angular momentum has relaxed toward the equilibrium state. \cite{inertial} 

Using $\dot{\vec \Omega } . \vec e_3 = 0$,  Eq.~(\ref{kinetic}), Eq.~(\ref{kineticBIS}) and the time derivative of Eq.~(\ref{kineticBIS}), Eq.~(\ref{Lagrange4}) re-writes:

\begin{equation}
\frac{d \vec M}{dt} = \gamma \vec M \times \left [ \vec H - 
\eta 
\left ( \frac{d \vec M}{dt} + \tau \frac{d^{2} \vec M}{dt^{2}} \right 
)  \right]
\label{Results}
\end{equation}

This is the Gilbert's equation of the dynamics of the magnetization that includes the new inertial term $- \gamma \tau  \, \vec M \times \left ( d^{2} \vec M/dt^{2} \right )$.

\section{Typical time scales}

 The limit of Eq.~(\ref{Results}) for $t \gg \tau$, where $\tau = \frac{I_1}{\eta 
M_s^2}$ leads to the kinetic limit. Since the damping $\eta$ can be replaced by the usual dimensionless Gilbert coefficient $\alpha = \gamma \eta M_s $, we have $\tau = \frac{I_1}{I_3} \frac{1}{\alpha \Omega_{3}}$. A rigorous study of the asymptotic behavior (as a function of the parameters $\gamma \vec H$, $\tau^{-1}$ and $\eta$) is beyond the scope of this work. However it is sufficient to observe that the limit $ \tau \rightarrow 0$ leads straightforwardly to the LLG equation:

 \begin{equation}
\frac{d \vec M}{dt} \, \longrightarrow  \, \gamma \vec M \times \left ( \vec H - 
\eta 
\frac{d \vec M}{dt} \right)
\label{Giblert2}
\end{equation}

In the same manner the vectorial gyromagnetic relation is recovered at 
the limit $\tau \rightarrow 0$. Eq.~(\ref{kinetic2}) gives:

\begin{equation}
\frac{d \vec L}{dt} = \eta \tau \,  \left ( \vec M \times 
\frac{d^2 \vec M}{dt^2} \right) + \frac{1}{\gamma} \frac{d \vec 
M}{dt} \, \longrightarrow \, \frac{1}{\gamma} \frac{d \vec M}{dt}
\end{equation}

The sufficient condition of validity of the Gilbert equation $I_1 = 0$ is hence
replaced by the condition $\tau \rightarrow 0$ (i.e. $ \tau \ll t$ for the relevant range of the parameters). 

An estimation of the value of $\tau$ gives the typical time scale for which inertial effects can be observed.  Here we 
come back to the simplest argument for the justification of the value of $\gamma$, namely the model of the Amp\`ere 
molecular currents. This is a quasi-one dimensional atomic model, for which the atomic orbital moment is defined by the electronic charge $q$ orbiting around a
nucleus at 
the distance $r$ with a velocity $v$. This system 
defines an electric loop that generates a magnetic moment 
$\vec M = \mathcal I S \, \vec e_z$, where $\mathcal I = q v/(2 \pi r)$ is the electric 
current, $S= \pi r^2$ is the surface enclosed by the loop, and $\vec 
e_z$ is the vector normal to the loop. This leads to the microscopic magnetic moment $M_s = q v r/2$. If we take the Bohr radius $a_0$ and the electron velocity $v$ with the Heisenberg relation $mv a_0 \ge \hbar/2$, the Bohr magneton is obtained for the minimum value of the atomic magnetic moment $ \mu_B = \gamma \hbar/2$. On the other hand, the angular moment of this system is 
$L_3= r mv$ and the ratio $M_3/L_3 \equiv \gamma = q/(2m)$. 
The angular frequency $\Omega_3$ is given by $L_3 = I_3~\Omega_3~=~\mu_B/\gamma$, i.e. $\Omega_3~=~\mu_B/(\gamma I_3)$ where $I_3 = m a_0^2$.   We have $\Omega_3 \approx 3 \, 10^{16} $ $rad / s $ and an order of magnitude of the typical times $\tau = \frac{I_1}{I_3} \frac{1}{\alpha \Omega_{3}}$ at which inertial effects should be observed is around a femtosecond (for a damping coefficient $\alpha$ such that $I_1 /(I_3 \,  \alpha) \approx 0.1$).

\section{Conclusion}
The paradoxical role played by the angular momentum for the dynamics of the magnetization has been studied in the light of the model introduced by T. H. Gilbert for the demonstration of the equation that bears his name. The demonstration has been reconsidered without the puzzling Gilbert's assumption of vanishing first moments of inertia $I_1=I_2=0$ and $I_3 \ne 0$. Instead, a general inertial tensor with the three arbitrary principal moments of inertia $ \{I_1, I_1,I_3\} $ has been used. A generalized expression of the equation of the dynamics of the magnetization is obtained, that includes an inertial term:  the mechanical analogy of the magnetic moment with the rigid rotator is complete. Both the usual expression of the Landau-Lifshitz-Gilbert equation and the usual gyromagnetic relation are recovered provided that a kinetic limit is performed for time scales much larger than the relaxation time of the angular momentum $\tau = I_1/(\eta M_s^2)$. The typical time scale is found to be of the order of the femtosecond.

\begin{references}

\bibitem{Stoehr} J. St\"ohr, H. C. Siegmann, {\it Magnetism. From Fundamental to Nanoscale Dynamics}, Springer, Berlin 2006.

\bibitem{Doering} This is not the case for non-uniform magnetization 
(domain walls or antiferromagnets), for which a magnetic mass is 
defined. See the pioneering work of W. D\"oring: "{\it \"Uber die 
tr\"agheit der W\"ande zwischen Weisschen Bezirken}" (On the inertia 
of walls between Weiss domains), Z. Natur. {\bf 3} 
373-379 (1948). D\"oring introduced the magnetic Lagrangian in this paper.

\bibitem{inertial} M.-C. Ciornei, J. M. Rub\'i, and J.-E. Wegrowe, {\it 
Magnetization dynamics in the inertial regime: Nutation predicted at 
short time scales}, Phys. Rev. B {\bf 83}, 020410(R) 1- 4 (2011).

\bibitem{Ohanian}  	H. C. Ohanian, {\it What is spin?}, Am. J. Phys. 54, 500 - 505 (1986)

\bibitem{Einstein}
The gyromagnetic relation ${\vec M} = \gamma {\vec L}$ has been 
established
through static magnetomechanical measurements, by S. J. Barnett (see 
Rev.  
Mod.  Phys.  {\bf 7}, 129 (1935)), and A.
Einstein and W. J. de Haas (Verh.  d. D. Phys.  Ges.  {\bf 17}, 152 
(1915)). 

\bibitem{Frenkel} V. Ya. Frenkel' {\it On the history of the 
Einstein-de Haas effect},  Sov. Phys. Usp. {\bf 22}, 580 - 587 
(1979).

\bibitem{Landau} L. Landau and E. Lifshitz,  {\it On the theory of 
dispersion of magnetic permeability in ferromagnetic bodies}, Phys. 
Z. Sowjet. {\bf  8}, 153-169 (1935).

\bibitem{Bloembergen} N. Bloembergen, {\it On the ferromagnetic 
resonance in Nickel and supermalloy}, Phys. Rev. {\bf 78}, 572-580 (1950)
	
\bibitem{GilbertPhD} T. L. Gilbert, {\it Formulation, foundations 
and applications of the phenomenological theory of ferromagnetism}, 
PhD dissertation, Illinois Institute of Technology, June 1956, 
appendix B.
	
\bibitem{Gilbert} T. L. Gilbert, {\it A phenomenological Theory of 
Damping in Ferromagnetic Materials}, IEEE Trans. Mag. {\bf 40}, 3443 
(2004). The discussion related to the assumption $I_1=I_2=0$ is 
confined in the footnotes 7 and 8 of the 2004 paper. Note that the 
original reference in Physical Review is only an abstract: T. L. Gilbert, {\it A Lagrangian formulation of the gyromagnetic 
equation of the magnetization fields}" Phys. Rev. {\bf 100}, 1243 
(1955).

\bibitem{Miltat} J. Miltat, G. Alburquerque, A. Thiaville, {\it An 
introduction to microsmagnetics in the dynamics regime}, in {\it 
Spin dynamics in confined magnetic structures I}, Eds. B. 
Hillebrands, K. Ounadjela, Springer, Berlin, 2002. The kinetic energy 
is introduced through the Lagrangian $\mathcal{L}$ page 19 Eq.~(37). 
The Lagrangian is such that (with our notations) $I_1 = I_2 = 0$ and 
$\Omega_3 = \dot \varphi \cos(\theta)$.


\bibitem{Brown} W. F. Brown Jr., {\it Single-Domain Particles : New 
Uses of Old Theorems}, Am. J. Phys. {\bf 28}, 542-551 (1960), see page 549.
        
  \bibitem{Morrish} A. H. Morrish, "{\it The Physical Principles of 
Magnetism}",  J. Wiley \& Son, New York 1965 (original edition), reprinted in 1980 by 
R. E. Krieger Publishing Company, and IEEE Press New York 2001. End of the page 551.
  
\bibitem{Ricci} T. F. Ricci and C. Scherer, {\it A Stochastic Model 
for the Dynamics of Classical Spin} , J. Stat. Phys. {\bf 67}, 1201 
(1992), page 1204-1208: "{\it In order to simulate the behavior of a 
classical spin, we take, for these equations, the limit $I_1 
\rightarrow 0$, $I_3 \rightarrow 0$, and $\dot \psi \rightarrow 
\infty$, but maintaining $I_3 \dot \psi = S(t) =$ finite }". In our 
notation $S  \equiv M_s$.

 \bibitem{Rubi} J. M. Rub\'i, A. P\'erez-Madrid, {\it Inertial effects in non-equilibrium thermodynamics}, Physica A {\bf 264} (1999) 492 - 502.
  
\bibitem{Condiff} D. W. Condiff and J. S. Dahler, {\it Brownian Motion of Polyatomic Molecules: The Coupling of Rotational and Translational Motions} 
J. Chem.  Phys.  {\bf 44}, 3988-4005 (1966).  

\bibitem{JEW} J.-E. Wegrowe "{\it Spin transfer from the point of view of the ferromagnetic degrees of freedom}", Solid State Com. {\bf 150}, 519- 523 (2010)

\bibitem{Rquedotpsi} The full calculation with $\dot 
\psi \ne 0$ gives the same result. It is consistent with that found with a different 
approach in reference M.-C. Ciornei {\it et al.} Phys. Rev. B {\bf 83}, 020410(R)
(2011).

\end{references}

\end{document}